\documentclass[pre,twocolumn, amsmath, amssymb, superscriptaddress]{revtex4}

\usepackage{graphicx}
\usepackage{dcolumn}
\usepackage{bm}
\usepackage{braket}
\usepackage{xcolor}

\usepackage{amsmath}

\newcommand{\beq}{\begin{equation}}
\newcommand{\eeq}{\end{equation}}

\begin{document}

\title{Investigating the side-chain structural organization behind the stability of protein folding and binding}




\author{Fausta Desantis}
\affiliation{Center for Life Nano \& Neuro Science, Istituto Italiano di Tecnologia, Viale Regina Elena 291,  00161, Rome, Italy}

\author{Mattia Miotto\footnote{Corresponding author: mattia.miotto@iit.it}}
\affiliation{Center for Life Nano \& Neuro Science, Istituto Italiano di Tecnologia, Viale Regina Elena 291,  00161, Rome, Italy}

\author{Lorenzo Di Rienzo}
\affiliation{Center for Life Nano \& Neuro Science, Istituto Italiano di Tecnologia, Viale Regina Elena 291,  00161, Rome, Italy}

\author{Edoardo Milanetti}
\affiliation{Department of Physics, Sapienza University, Piazzale Aldo Moro 5, 00185, Rome, Italy}
\affiliation{Center for Life Nano \& Neuro Science, Istituto Italiano di Tecnologia, Viale Regina Elena 291,  00161, Rome, Italy}

\author{Giancarlo Ruocco}
\affiliation{Center for Life Nano \& Neuro Science, Istituto Italiano di Tecnologia, Viale Regina Elena 291,  00161, Rome, Italy}

\affiliation{Department of Physics, Sapienza University, Piazzale Aldo Moro 5, 00185, Rome, Italy}

\begin{abstract}
What are the molecular mechanisms that dictate protein-protein binding stability and whether those are related to the ones behind protein fold stability are still largely open questions. Indeed, despite many past efforts, we still lack definitive models to account for experimental quantities like protein melting temperature or complex binding affinity.
Here, we investigate and compare chemical and physical features on a dataset of protein with known melting temperature as well as a large dataset of protein-protein complexes with reliable experimental binding affinity. In particular, we probed the aminoacid composition and the organization of the network of intramolecular and intermolecular  interaction energies among residues. 

We found that hydrophobic residues present on the protein surfaces are preferentially located in the  binding regions, while charged residues behave oppositely. In addition, the abundance of polar amino acid like Serine and Proline correlates with the binding affinity of the complexes. 
Analysing the interaction energies we found that distant Coulombic interactions are responsible for thermal stability while the total inter-molecular van der Waals energy correlates with protein-protein binding affinity. 
\end{abstract}

\maketitle

\section{Introduction}

Interactions between biomolecules are at the basis of every cellular processes, from DNA replication to protein degradation \cite{keskin2008principles,nooren2003structural,perkins2010transient}.
It is important to note that, in most cases, to ensure a proper molecular recognition between proteins, they must be appropriately folded \cite{jones1996principles, gromiha2017protein}.
From this point of view, the structural and energetic properties of protein folding and binding are interrelated \cite{manhart2015protein}.

In both the case of protein folding and protein-protein interaction indeed, the spatial arrangements of residues side chains give rise to a complex network of non-bonded atom-atom interactions, whose characteristics are important for structure stability. Indeed, the non-bonded intra-molecular interactions in proteins, both in terms of their spatial and energetic arrangements, play a key role on the thermal stability of the protein structure \cite{folch2010thermo,miotto2019insights, Miotto2020}. Similarly, protein-protein inter-molecular interactions are surely important for the binding affinity between the two interacting molecules \cite{Zhang_2011,10.1371/journal.pone.0110085,vangone2015contacts,BROCK20073340,Miotto2021}.

In this scenario, the energetic of non-bonded atomic interactions can be summarized mainly by Coulomb and Van der Waals forces.
To quantitatively analyze folding properties and binding properties, two descriptors are usually adopted: (i) the thermal resistance of each protein is typically evaluated through the melting temperature ($T_m$) \cite{folch2010thermo}, while (ii) the affinity of the interaction between two proteins is described through the equilibrium dissociation constant ($K_d$) \cite{vangone2015contacts}.

Predicting the stability of a protein through a structure-based approach is still an open challenge, since a complete understanding of the relationship between thermal resistance and the reorganization of the internal energies of the protein is still lacking \cite{folch2010thermo}. In this context, some recent studies reported differences, in terms of amino acid composition or spatial arrangement of residues, characterizing pairs of thermostable-mesostable homologs proteins \cite{amadei2018density, vijayabaskar2010interaction}.
Moreover, even if core packing seems related to thermal resistance at least to some extent \cite{vogt1997protein}, it has been demonstrated that the residues hydrophobicity play a rather marginal role on protein stabilization \cite{priyakumar2012role, van1994protein}: the pivotal role seems to be played by electrostatic residues on the protein surface \cite{miotto2019insights}.


Similarly, many computational methods have been developed to try to understand and, possibly, predict the basic mechanisms of the binding affinity between molecules \cite{vangone2015contacts,qin2011automated,audie2007novel}. For example, it was found that the presence of ALA anti-correlates with affinity suggesting that such a residue will not provide favorable interactions\cite{10.1371/journal.pone.0110085}.
However the comprehension of the structural determinants of protein-protein binding is still far from being achieved. It is testified by the difficulties of predictive methods, especially those based on empirical functions typically used in molecular docking\cite{jiang2002potential,ma2002fast,luo2014functional}, to successfully reproduce the results, especially when a large dataset is considered \cite{fleishman2011community}.

To this end, given the functional interconnection between the folding and binding processes, here we present a computational study to compare the distribution of the interaction energies of folding (intra) and binding (inter). For this purpose, we consider both the Coulomb and van der Waals energies to investigate their relationship with thermal resistance and binding affinity, considering intramolecular and intermolecular interactions respectively. In fact, we consider two datasets of experimentally resolved molecular structures, the first composed of single chain proteins for which the melting temperature ($T_m$) value is known\cite{miotto2019insights} and the second one composed of experimental protein-protein complexes for which the experimental value of binding affinity is reported\cite{dias}.

We preliminary studied the frequencies of different amino acids in such proteins, highlighting as different fold stability or binding properties are reflected in the preferred occurrence of residues with different chemical characteristics. 

Thus, the predominant role of the Coulomb term in intramolecular interactions is shown, underlining the role of this contribution for thermal stability, as previously shown in \cite{zhou_thermo}. By analogy, we show the opposite behavior of the contribution of the Coulomb term in relation to the binding affinity between two interacting molecules.
A particular focus is addressed on the role of van der Waals energy, distinguishing its energy distribution in intramolecular or intermolecular interactions. In the latter case, in fact, we show the relationship between the van der Waals interactions and the binding affinity, also focusing on the connection between the van der Waals description of the molecular interfaces and the corresponding shape complementarity between the two interacting molecular surfaces.\\

\section{Results and Discussions}

\subsection{The role of the amino acid composition in intra- and intermolecular interactions}

In order to investigate how the amino acid composition plays a role in both folding and binding properties, we analyze the frequency of occurrence of each amino acid in the protein structures of the two datasets considered in this comparative study, i.e. the Thermostability  dataset, composed of single-chain proteins with the known experimental melting temperature ($T_m$); and  the Affinity dataset, which is composed of molecular complexes with a known experimental structure with the corresponding value of binding affinity (see Methods for details).

\begin{figure*}[t]
\centering
\includegraphics[width = 0.8\textwidth]{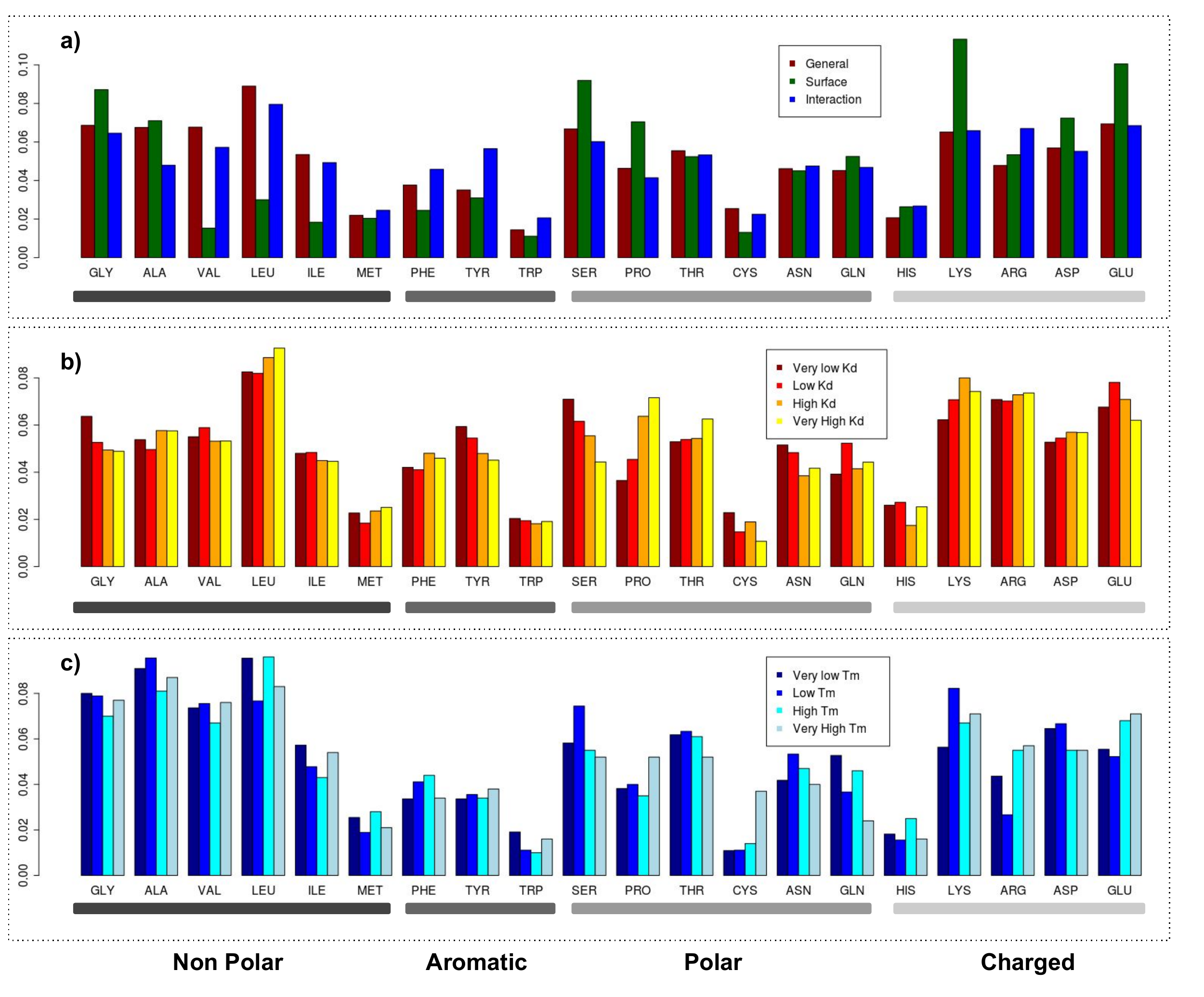}
\caption{\textbf{Comparison between the amino acid composition of proteins with different thermal stability and complexes with different binding affinity.} \textbf{a)} Relative abundances of each of the twenty natural aminoacid in the Thermostability and Affinity datasets. For each kind of amino acid, green bar corresponds to the abundance found in the solvent exposed residues; blue one to the residues found in interaction with the molecular partner and red bars are computed considering all residues. \textbf{b)} Relative abundances of each of the twenty natural aminoacid found in the binding site regions of the  Affinity dataset stratified by four groups of different binding  affinities.  Bar colors range from dark red to yellow as the $k_d$ of the considered complexes increases. \textbf{c)} Relative abundances of each of the twenty natural aminoacid found in  the  Thermostability dataset stratified by four groups according to the protein melting temperatures ($T_m$).  Bar colors range from dark blue to light blue as the thermal stability of the considered proteins increases.}
\label{fig:1}
\end{figure*}

The results of this analysis is reported in Figure \ref{fig:1}. In panel a) we have shown a general overview using all the proteins in the 2 dataset. In red the overall frequencies of residues occurrence in protein are reported. In green we recorded the values restricting to solvent-exposed residues (see Methods for the definition of superficial residue). In blue we shown the frequencies of the amino acid obseved to be in interaction in the Affinity dataset, where a residue is considered to be in contact if it has at least one atom closer than 4 $\AA$ to its molecular partner.

As well known, we found that hydrophobic amino acids, such as VAL or LEU or ILE, are poorly present in the solvent-exposed surface of proteins. However, when a hydrophobic amino acid is present in the exposed regions it has a high probability of interacting with the corresponding molecular partner. On the contrary, charged amino acids, such as LYS or GLU, are typically very present in the superficial protein region. This notwithstanding, the fraction of these actually in interaction is relatively small, probably because the function of these superficial charged residues in linked with fold stability and not with partner recognition.

In addition, in Figure \ref{fig:1}.b) we studied the frequencies of amino acids found in protein binding sites, separated according to the binding affinity with partners. Each bar color represent a quartile of the distribution, meaning that the dark red and light yellow bars regard the 25\% of protein-protein complexes in our dataset with the lowest or highest $K_d$, respectively. It is interesting to note the peculiar behavior of some residues. For example it seems that an high presence of PRO is characteristic of high $K_d$ complexes, while SER and TYR seems to favorite stroger associations.

Lastly, in Figure \ref{fig:1}.c) we investigated the amino acids frequencies in proteins characterized by different $T_m$. As in the previous plot, each bar represent a quartile of the $T_m$ distribution, from dark blue (very low $T_m$) to light blue (very high $T_m$). Interestingly, we found that an high presence of CYS is typical of proteins with very high $T_m$, since such a residue is responsible for the formation of stabilizing disulfide bridges. Moreover the presence of a high number of charged residues, such as ARG or GLU, seems to be correlated with an higher $T_m$.

\subsection{Comparison between interaction energy terms for thermal stability and binding affinity properties}

The non-bonded interactions between the residue side chains are the basis both for protein folding and for protein-protein binding \cite{rao2014protein}, contributing to structural stabilization \cite{chakrabarty2016naps}. In particular, for this study we considered only the residue-residue interactions neglecting the protein-solvent interactions since the inclusion of these interactions would require a dynamic approach \cite{timasheff1970protein}.

\begin{figure*}[t]
\centering
\includegraphics[width = 0.8\textwidth]{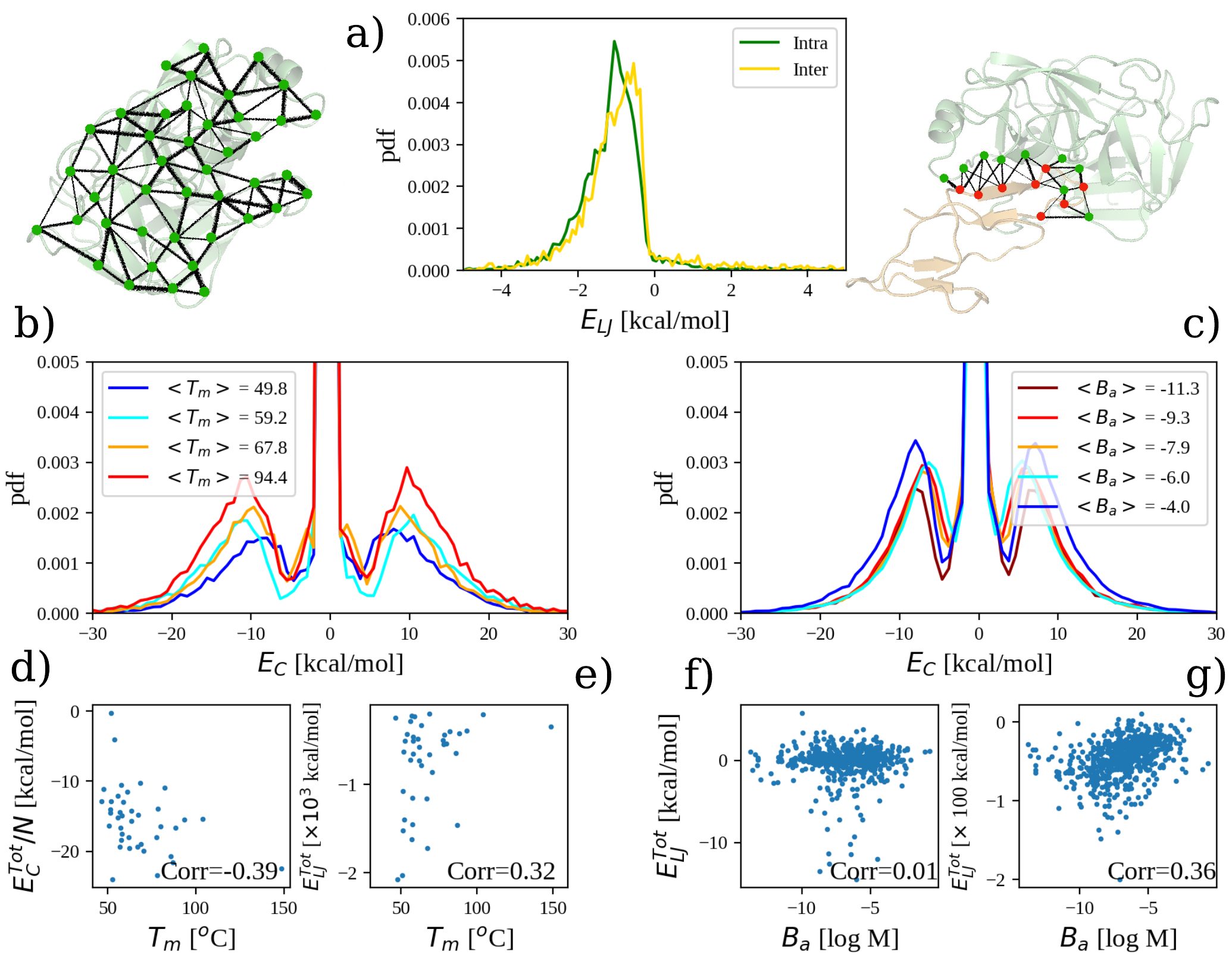}
\caption{\textbf{a)} Probability density distributions of Lennard-Jones potential energy for the Thermostability dataset ('Intra',green line) and between the each couple of proteins of the Affinity dataset ('Inter', yellow line). Energies are considered only between couples of residues whose minimum distance is lower than 4 $\AA$ (see Methods). 
\textbf{b)} Probability density distributions of Coulombic interaction energies for the four subsets defined in the Thermostability dataset from lower (dark blue) to higher (dark red) $T_m$. Each distribution is built using a group of proteins whose melting temperatures lie in the same range; the average $T_m$ value of each group is reported in the legend. The density functions exhibit a dependence with the melting temperatures ranges and peak heights increase with the temperatures.
\textbf{c)} Probability density distributions of Coulombic interaction energies for the four subsets defined in the Affinity dataset from lower (dark red) to higher (dark blue) $K_d$. Each distribution is built using a group of protein complexes whose binding affinities lie in the same range; the average $log_{10}(K_d)$ value of each group is reported in the legend. The density functions exhibit a dependence with the binding affinity ranges and peak heights decreases with the affinity.  \textbf{d)}  Correlation between $T_m$ and the total Coulombic energy of each protein of the Themostability dataset normalized by the protein size. \textbf{e)}  Same as in d) but for the total Lennard Jones potential energy.
\textbf{f)}  Correlation between $K_m$ and the total Coulombic energy of each protein of the Affinity dataset normalized by the protein size. \textbf{g)}  Same as in f) but for the total Lennard Jones potential energy.
}
\label{fig:2}
\end{figure*}

Here we present a computational analysis of how the energetic contributions of Coulomb and van der Waals affect the thermal stability of protein structures and the binding affinity in molecular complexes. 
At first, we analyze Lennard-Jones (LJ) potential energy both for intramolecular interactions (responsible for folding) and for intermolecular interactions (responsible for binding between two molecules). In Figure~\ref{fig:2}a, we show the distribution of both LJ interactions intramolecular (in green) and intermolecular (in yellow). The two curves are characterized by similar trends, where most of the area under the curve corresponds to negative (favorable) energy values. Therefore, both types of interactions show how the side chains optimize their spatial rearrangement to minimize the energetic contribution. However, a difference between the two curves can be noted: interestingly, the LJ probability density of intermolecular interactions is shifted towards positive values with respect to the intramolecular interactions curve, indicating that the LJ interactions that characterize the fold of a protein tend to be more negative than the intermolecular interactions.
A comparative analysis of the LJ interaction probability densities for several subgroups of different Tm and for several subgroups of different binding affinity, highlights that there is no significant discrepancy between the various curves (data not shown).

Secondly, we investigated the relationship between thermostability and Coulomb energy distribution of intramolecular interactions, as well as the relationship between binding affinity and Coulomb energy distribution of intermolecular interactions (see Figure~\ref{fig:2}). To this end, the Tm dataset was divided into 4 groups according to protein Tm and for each group the energy distribution was evaluated. Similarly, the affinity dataset was divided into 5 groups according to the binding affinity experimental values.
The general shape of the density functions is almost identical between the 4 cases, independently from the thermal properties of the macromolecules, and this is clearly due to the general folding energetic requirements.

A strong dependence between thermal stability and the percentage of strong interactions is evident looking at the disposition of the density curves (Fig.~\ref{fig:2}b): the higher the thermal stability the higher the probability of finding strong interactions. On the contrary, less thermostable proteins possess a larger number of weak interactions. 
Similarly, the same analysis was performed to investigate the relationship between the probability density distributions of Coulomb energies and the experimental binding affinity of the analyzed complexes. To this end, we evaluate the Coulombic potential energies of the 620 complexes of the Affinity dataset\cite{dias}, splitting the dataset into 5 groups corresponding to different ranges of affinity values.


The analysis shows that the inter- and intermolecular interactions are characterized by the same distribution of Coulomb energy, since the shapes of all the curves are comparable (see Figure~\ref{fig:2}). In particular, both intermolecular and intramolecular coulombic interactions are characterized by two well discernible peaks being symmetric with respect to zero (discarding the less informative zero-centered peak) centered at $E_C \sim\pm 7 kcal/mol$. These peaks account for favourable (negative peak) and unfavourable (positive peak) interactions. 
Their symmetry means that the two balance each other, contrary to what happens for the van der Waals energy distributions. 
The comparative analysis between the different curves of the probability density of Coulombian energies for different ranges of binding affinity, shows a trend opposite to the trend of the curves obtained by dividing the proteins by range of melting temperatures. 
Indeed, as shown in Figure~\ref{fig:2}c, the higher the binding affinity, the lower the maximum of the distribution peak in the range of strong interactions (peaks centered at about $7~kcal/mol$  and $-7~kcal/mol$).

This behavior could be interpreted as a counterintuitive result as a less stable condition seems to be more likely compared to cases in which the affinity is higher and related to more stable complexes. On the other hand, the presence of two symmetric peaks with respect to zero leads to the hypothesis that repulsive and attractive interactions counterbalance each other thus explaining why the global result is a poor binding driving force represented by the higher values of $K_d$.\\ The existence of these symmetric peaks could be explained with the fact that protein interfaces host atoms of both charges on both sides so that either repulsive and attractive interactions are possible. In this view, it could be stated that complexes characterized by a lower $K_d$ are those characterized by the right number and arrangement of charged residues with respect to complexes having larger $K_d$ values which are probably too rich in charged residues instead.

Finally, we perform a correlation analysis between the inter- and intramolecular energies and the experimental values of thermal stability and binding affinity. In particular, we consider the average Coulomb energy for each residue, since the Coulomb term can have long range contributions and therefore depend on the size of the molecular systems. On the contrary, only the total sum of the energies is taken into account for the quantification of the interaction of the Lennard-Jones potential. 
The analysis shows a direct negative linear correlation between the experimental thermal stability and the average Coulomb energy of each protein, in accordance with the previous analysis on the distributions of the energy probability densities. More specifically, the Pearson's correlation value is -0.39 with a p-value statistically significant of 0.013. Even more interestingly, a highly significant linear dependence exists between the total energy of LJ potential and the experimental binding affinity (considered in log basis). In this case the pearson correlation is 0.36 with a p-value lower than $10^{-4}$. Therefore, this linear relationship is much more robust than that between the sum of the van der Waals intramolecular interactions and the $T_m$ value of each protein.
No correlation exists between the mean of the Coulombian intermolecular interactions and the experimental binding affinity values (Pearson correlation of 0.01, with a p-value of 0.80).

\subsection{Exploring energy organization}
The significant correlation between the sum of the energies of the Lennard-Jones potential and the experimental binding affinity invites us to investigate the role of the LJ energies at a higher level of description. In fact, in order to analyze the energy reorganization of binding, we calculate the sum of the energy co-attributes between each residue and all its interacting residues.

To this end, thinking of residues in a protein as nodes in a network, we can define the node strength as~\cite{miotto2019insights}:
\begin{equation}
s_i = \sum_{j = 1}^{N_{aa}^i} E_{ij}
\end{equation}

where $N_{aa}^i$ is the number of residues found in interaction with  residue $i$.\\ 

\begin{figure*}[t]
\centering
\includegraphics[width = 0.9\textwidth]{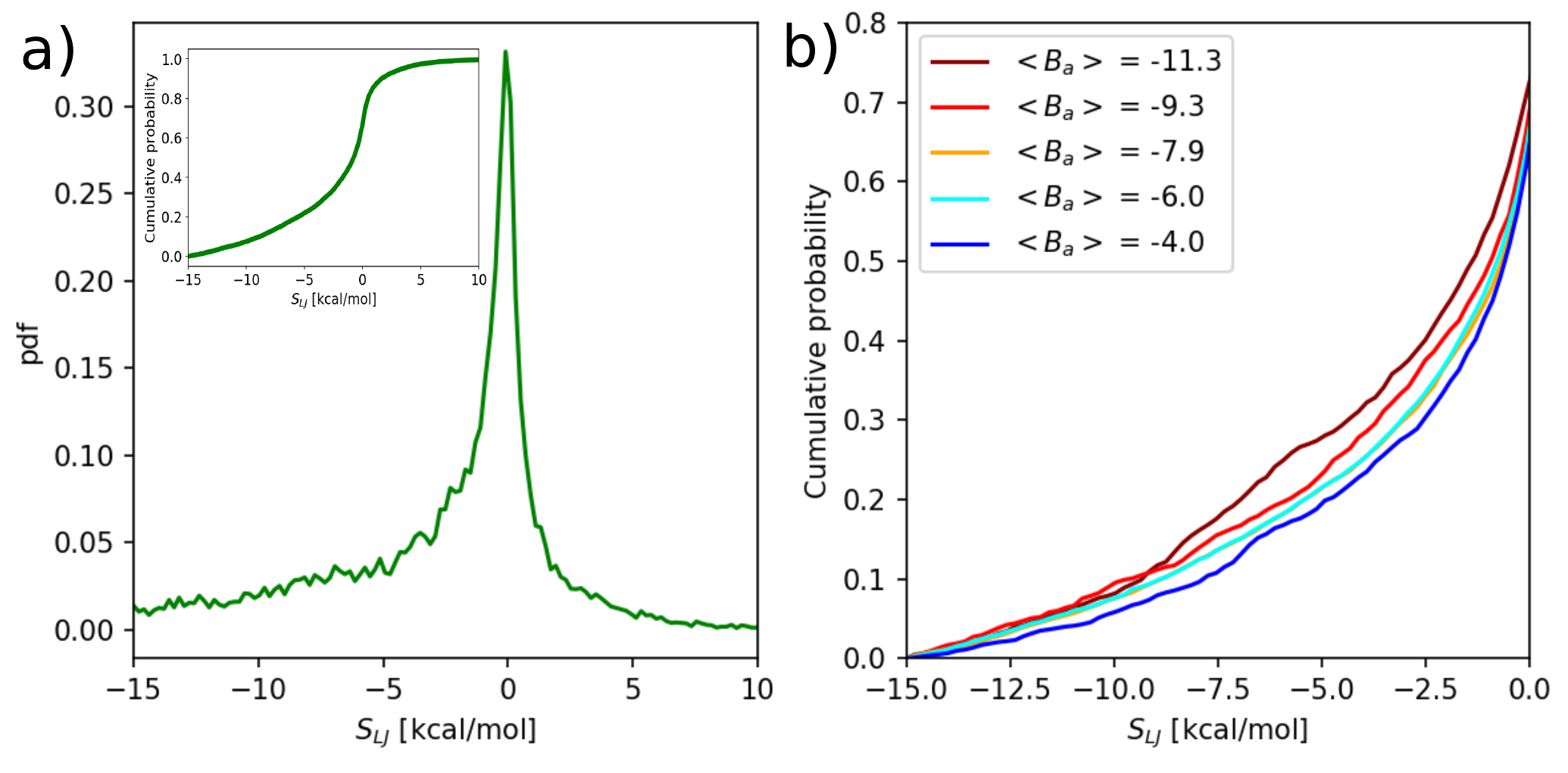}
\caption{\textbf{a)} Probability distribution of the Strength values obtained using the Lennard Jones intermolecular energies on the 622 complexes of the Affinity dataset. Comulative distribution is shown in the inset. \textbf{b)} Zoom of the cumulative distributions in the negative Strength range obtained dividing the Affinity dataset in five groups according to the binding affinity, $B_a$, of the complexes. 
}
\label{fig:3}
\end{figure*}

Figure~\ref{fig:3}a shows the probability density of the van der Waals strengths at the interface (i.e. considering only intermolecular interactions).  It can be seen  a marked peak around zero, meaning that the low intensity LJ interactions (in most cases the farther away in terms of spatial distance) are more likely. However, the behavior of the curve for the two tails is interesting, considering both the direction of positive and negative energies. In accordance with the formation of the molecular complex, the negative LJ interactions are more organized, since it is more likely to find a residue whose sum of the van der Waals energies is strongly negative compared to the case of finding a residue that interacts with each other with strong positive energies (see Cumulative distribution in the inset).

Interestingly, as found in  the case of thermal resistance~\cite{miotto2019insights}, the organization of the energies, measured by the network strength parameter, shows a different behaviour when looking at complexes with different binding affinity. Indeed, looking at Cumulative distributions of the LJ strength divided according to the complex binding affinity (see Figure~\ref{fig:3}b), the higher the probability of finding a residue with high favorable energy, the higher the complex binding affinity.
All together, these analyses identify van der Waals interactions as a key component for the complex stability.

The property of van der Waals interactions to act on a short range allows to hypothesize that at the interface protein atoms may be arranged in such a way that protein surfaces are compatible. This fact suggests that shape complementarity is a reasonable parameter to describe binding.

\section{Conclusions}
The analysis of the energy distribution of intra and intermolecular interactions is fundamental to shed light on the molecular mechanisms that determine both the thermal stability of the protein structures and the binding affinity between two interacting molecules, respectively.
For this purpose, here we propose a comparative computational analysis between non-bonded interactions responsible both for folding properties and binding properties. In fact, two datasets with known experimental data of melting temperature and binding affinity were selected to study the influence of the energy/structural organization on the two properties considered.
Our work shows the importance of the role of Coulomb interactions for the thermal stability of proteins, as well as the importance of Lennard-Jones interactions for the characterization of the binding affinity for molecular complexes. Furthermore, a linear anti-correlation trend was found between binding affinity and the distribution of strong intramolecular Coulomb energies, indicating the role of the spatial arrangement of charged amino acids in solvent-exposed regions.

\section{Methods}

\subsection{Datasets}

In order to compare the residue organization and composition between protein with stable structures and complexes with stable binding, we collected two datasets:  
\begin{itemize}
    \item The Thermostability ($T_m$) dataset,  which consisted of 40 single-chain structures with known melting temperature, $T_m$~\cite{miotto2019insights}. 
\item The Affinity ($B_a$) dataset was obtained from \cite{dias} and consisted of 620 complexes of known dissociation constant. Namely, the dissociation constant $K_d$, the inhibition constant $K_i$ and the half maximal inhibitory concentration $IC_{50}$ were reported. 
\end{itemize}

In order to remove structural alterations (missing atoms, crystallographic clashes etc.), all protein structures were treated with PDBFixer\cite{Eastman2013a}  to replace or remove non standard residues (heterogens), and to add any missing heavy atoms.
After that, all proteins of the Thermostability datset and all complexes of the Affinity dataset were minimized using the standard GROMACS algorithm and the
CHARMM force field~\cite{charmm,GRO} in vacuum. Minimization was carried out using a steepest descent algorithm arresting the simulation when the maximum force was less than $1kJ/mol$.

\subsection{Interaction energy calculation}

Intra and inter-molecular interaction energies were computed using the parameters obtained from CHARMM force-field. In particular, given two atoms $a_l$ and $a_m$ holding partial charges $q_l$ and $q_m$ , the Coulombic interaction between them can be computed as:
\begin{equation}
\small
\label{eq:C}
E_{lm}^C= \frac{1}{4\pi\epsilon_0}\frac{q_lq_m}{r_{lm}}
\end{equation}
where $r_{lm}$ is the distance between the two atoms, and $\epsilon_0$ is the vacuum permittivity. Van der Waals interactions can instead be calculated as a 12-6 Lennard-Jones potential:
\begin{equation}
\small
\label{eq:LJ}
E_{lm}^{LJ} = \sqrt{\epsilon_l \epsilon_m}\left[ \left(\frac{R_{min}^l + R_{min}^m}{r_{lm}}\right)^{12} - 2\left(\frac{R_{min}^l + R_{min}^m}{r_{lm}}\right)^{6}\right]
\end{equation}
where $\epsilon_l$ and $\epsilon_m$ are the depths of the potential wells of $a_l$ and $a_m$ respectively, $R_{min}^l$  and $R_{min}^m$ are the distances at which the potentials reach their minima.

The total interaction energy  between each couple of residues is defined as:

\begin{equation}
    E_{AA_{ij}}^X= \sum_{l = 1}^{N_{atom}^i }\sum_{m = 1}^{N_{atom}^j} E_{lm}^X
\end{equation}

where $E_{AA_{ij}}^X$ is the energy between two amino acids $i$ and $j$, obtained as the sum of the interactions between each atom of the two residues ($N_{atom}^i$, $N_{atom}^j$); $X$ stands for the kind of interaction considered, either Coulombic ($X = C$) or Lennard-Jones ($X = LJ$).

As for the distance between a pair of residues, this was assessed selecting the minimum distance between the atoms composing them.


\end{document}